\documentclass[%
 aip,
 jcp,
 amsmath,amssymb,
 reprint,%
]{revtex4-1}

\usepackage{graphicx}
\usepackage{dcolumn}
\usepackage{bm}

\draft 

\begin{document}

\preprint{AIP/123-QED}

\title{Propagation of hydrodynamic interactions between particles in a compressible fluid} 



\author{Rei Tatsumi}
 \email{tatsumi@cheme.kyoto-u.ac.jp}
\author{Ryoichi Yamamoto}%
 \email{ryoichi@cheme.kyoto-u.ac.jp}
\affiliation{%
 Department of Chemical Engineering, Kyoto University, Kyoto 615-8510, Japan
}%


\date{\today}

\begin{abstract}
Hydrodynamic interactions are transmitted by viscous diffusion and sound propagation, and the temporal evolution of hydrodynamic interactions by both mechanisms is studied 
using direct numerical simulation in this paper.
The hydrodynamic interactions for a system of two particles in a fluid 
are estimated using the velocity correlation of the particles.
In an incompressible fluid, hydrodynamic interactions propagate instantaneously at the infinite speed of sound followed by 
a temporal evolution due to viscous diffusion.
Conversely, sound propagates in a compressible fluid at a finite speed, which affects 
the temporal evolution of the hydrodynamic interactions 
through an order-of-magnitude relationship between the time scales of viscous diffusion and sound propagation.
The hydrodynamic interactions are characterized 
by introducing the ratio of these time scales as an interactive compressibility factor.
\end{abstract}

\pacs{Valid PACS appear here}

\maketitle 


\section{\label{sec:level1}Introduction}
In particle dispersions, the motion of each particle in a fluid solvent affects the motion of the other particles.
Such dynamical interactions are called hydrodynamic interactions; these interactions  
produce complex dynamical behavior for dispersions that can be observed in particle aggregation and sedimentation phenomena.

Hydrodynamic interactions correspond to momentum exchange among particles through the ambient fluid 
and are transmitted by two mechanisms: viscous diffusion and sound propagation.
These two mechanisms occur at different time scales.
The time scale for viscous diffusion over a distance equal to the particle size is $\tau_\nu = a^2/\nu$, 
while the time scale for sound propagation is $\tau_c = a/c$, 
where $a$ is the particle radius, $\nu$ is the kinematic viscosity, and $c$ is the speed of sound in the fluid.
To study dynamical effects at the scale of particle size, the relative significance of the sound propagation mechanism 
in hydrodynamic interactions is assessed from the ratio of the two time scales described above:
\begin{eqnarray}
\varepsilon = \frac{\tau_c}{\tau_\nu} = \frac{\nu}{ac}
\label{e3-1-1}.
\end{eqnarray}
This dimensionless quantity is called the compressibility factor.
Because sound propagation is much faster than viscous diffusion, the compressibility factor is generally quite small;
for example, in a dispersion of water and particles of radius $a = 100\:{\rm nm}$, 
the compressibility factor is estimated to be $\varepsilon \approx 7 \times 10^{-3}$.
In this case, the assumption of incompressibility is fully justified.
Thus, fluids are assumed to be incompressible in many theoretical studies of hydrodynamic interactions,
and only viscous diffusion is considered as the temporal evolution mechanism for hydrodynamic interactions ~\cite{B3-1, B3-2}.
However, the speed of sound in a liquid is around $10^3 \ {\rm m/s}$ irrespective of the liquid considered, and this factor can produce a large 
compressibility factor for a dispersion of a highly viscous fluid solvent, 
e.g., $\varepsilon \approx 0.7$ for olive oil and $\varepsilon \approx 10$ for corn syrup.

In recent years, the temporal evolution of hydrodynamic interactions between two particles has been directly observed~\cite{B3-3, B3-4, B3-5, B3-6}.
In these experimental studies, particles were trapped by optical tweezers, and the correlations between the positional fluctuations of particles were measured.
The authors reported the temporal evolution of hydrodynamic interactions in the viscous diffusion regime,
which coincided with analytical predictions that assumed the fluid was incompressible.
Evidence for hydrodynamic interactions caused by sound propagation was also observed~\cite{B3-3, B3-7, B3-8} 
but could not be captured due to the extremely short time-scale of sound propagation.

In the present study, we investigate the propagation process of hydrodynamic interactions using direct numerical simulation.
Within this approach, the hydrodynamic interactions are directly computed by simultaneously solving for 
the motions of the fluid and the particles with appropriate boundary conditions. 
We use the smoothed profile method (SPM)~\cite{B3-9, B3-10},
which can be applied to a compressible fluid as well as an incompressible fluid~\cite{B3-11}.
By considering the fluid compressibility, sound propagation can be captured.

We consider a system of two particles in a fluid and investigate the correlated motion of the particles.
In particular, we estimate the velocity relaxation of one particle in response to the exertion of an impulsive force on the other particle.
This velocity cross-relaxation function is equivalent to the velocity cross-correlation function 
from the fluctuation-dissipation theorem.
The velocity cross-relaxation function may be interpreted 
in terms of the temporal evolution of the flow field around the particle.
We first consider an incompressible fluid to identify the separate characteristics of sound propagation and viscous diffusion in the hydrodynamic interactions.
We then consider a compressible fluid to investigate the effect of the order-of-magnitude relationship between the
sound propagation and viscous diffusion time scales on the temporal evolution of the hydrodynamic interactions.
In addition, we examine the validity of analytical solutions within the Oseen approximation, which are often compared with experimental results.

\section{Model}

\subsection{Basic equations}

We model a dispersion as a system in which spherical particles are dispersed in a Newtonian fluid.
The motion of the particles is governed by Newton's and Euler's equations of motion, which can be written for the $i$-th particle as
\begin{eqnarray}
M_i \frac{\mathrm{d}}{\mathrm{d} t} \boldsymbol{V}_i = \boldsymbol{F}^H_i + \boldsymbol{F}^C_i + \boldsymbol{F}^E_i,\ \ \ \ 
\frac{\mathrm{d}}{\mathrm{d} t} \boldsymbol{R}_i = \boldsymbol{V}_i
\label{e3-2A-1},
\end{eqnarray}
\begin{eqnarray}
\boldsymbol{I}_i \cdot \frac{\mathrm{d}}{\mathrm{d} t} \boldsymbol{\Omega}_i = \boldsymbol{N}^H_i + \boldsymbol{N}^E_i
\label{e3-2A-2},
\end{eqnarray}
where $\boldsymbol{R}_i$, $\boldsymbol{V}_i$, and $\boldsymbol{\Omega}_i$ are the position, the translational velocity, 
and the rotational velocity of the $i\mathchar`-$th particle, respectively.
The particle has a mass $M_i$ and a moment of inertia $\boldsymbol{I}_i$.
The fluid exerts a hydrodynamic force $\boldsymbol{F}^H_i$ and a torque $\boldsymbol{N}_i^H$ on the particle, while
a force $\boldsymbol{F}^C_i$ is exerted through direct interactions among the particles.
A force $\boldsymbol{F}^E_i$ and a torque $\boldsymbol{N}^E_i$ are externally applied.
The hydrodynamic force and torque are evaluated by considering the momentum conservation between the particle and the fluid.

The fluid dynamics are governed by the following hydrodynamic equations:
\begin{eqnarray}
\frac{\partial \rho}{\partial t} + \boldsymbol{\nabla} \cdot (\rho \boldsymbol{v}) = 0
\label{e3-2A-3},
\end{eqnarray}
\begin{eqnarray}
\frac{\partial \rho \boldsymbol{v}}{\partial t} + \boldsymbol{\nabla} \cdot (\rho \boldsymbol{vv}) 
= \boldsymbol{\nabla} \cdot \boldsymbol{\sigma} + \boldsymbol{f}^R
\label{e3-2A-4},
\end{eqnarray}
where $\rho(\boldsymbol{r},t)$ and $\boldsymbol{v}(\boldsymbol{r},t)$ are the mass density and velocity fields, respectively, of the fluid.
The stress tensor is given by
\begin{eqnarray}
\boldsymbol{\sigma} = -p \boldsymbol{I} + \eta [\boldsymbol{\nabla} \boldsymbol{v} + (\boldsymbol{\nabla} \boldsymbol{v})^T]
 + \left( \eta_v - \frac{2}{3}\eta \right)(\boldsymbol{\nabla} \cdot \boldsymbol{v}) \boldsymbol{I} \nonumber, \\
\label{e3-2A-5}
\end{eqnarray}
where $p(\boldsymbol{r},t)$ is the pressure, $\eta$ is the shear viscosity, and $\eta_v$ is the bulk viscosity.
A body force $\boldsymbol{f}^R (\boldsymbol{r},t)$ is added to satisfy particle rigidity.
We also assume a barotropic fluid described by $p = p(\rho)$
with a constant speed of sound $c$ such that
\begin{eqnarray}
\frac{\mathrm{d} p}{\mathrm{d} \rho} = c^2
\label{e3-2A-6}.
\end{eqnarray}
Equations~(\ref{e3-2A-3})-(\ref{e3-2A-6}) are closed for the variables $\rho$, $\boldsymbol{v}$, and $p$ 
without considering energy conservation.

We use the SPM to perform direct numerical simulations.
In this method, the boundaries between particle and fluid are modeled by a continuous interface.
For this purpose, a smoothed profile function $\phi(\boldsymbol{r}, t) \in [0, 1]$ is introduced 
to distinguish between the particle and fluid domains,
i.e., $\phi = 1$ in the particle domain and $\phi = 0$ in the fluid domain.
These two domains are smoothly connected through a thin interfacial region of thickness $\xi$.
The body force for the particle rigidity $\boldsymbol{f}^R$ is expressed as $\rho \phi \boldsymbol{f}_p$.
The detailed mathematical expressions of $\phi$ and $\phi \boldsymbol{f}_p$ have been previously given~\cite{B3-9}.

\subsection{Linear formulation}

\begin{figure}[tbp]
\centering
\includegraphics[width=80mm, clip]{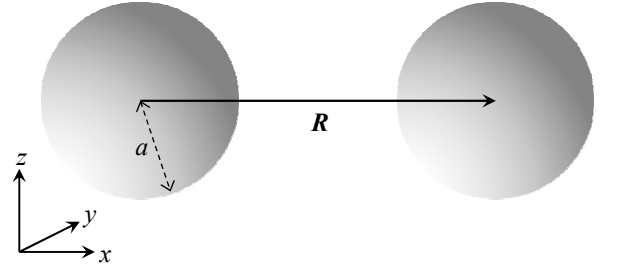}
\caption{\label{f3-1} Geometry of the present model system.
Two spherical particles of equal radius $a$ are situated in a fluid with a center-to-center distance $R = |\boldsymbol{R}|$.
This system is axisymmetric about the $\boldsymbol{R} = \boldsymbol{R}_2 - \boldsymbol{R}_1$ direction 
such that two directions are of interest: parallel and perpendicular to the center-to-center axis of the particles.
In this figure, the $y$ and $z$ directions are degenerate perpendicular directions.
}
\end{figure}

The hydrodynamic equations can be linearized for low Reynolds number flows in a dispersion.
Correspondingly, the equations of motion of the particles are also linear.
Most approximate theories are based on this formulation.

Here, for an $N_p$-particle system, we define the following column vectors, which are $6N_p$-dimensional vectors, to describe the equations concisely:
\begin{eqnarray}
\boldsymbol{U}
= \left( \begin{array}{l} 
\boldsymbol{V}_1 \\ \ \vdots \\ \boldsymbol{V}_{N_p} \\[0.4em] \boldsymbol{\Omega}_1 \\ \ \vdots \\ \boldsymbol{\Omega}_{N_p} \\ 
\end{array} \hspace{-0.4em} \right), \ \ \
\boldsymbol{H}
= \left( \begin{array}{c} 
\boldsymbol{F}^H_1\\ \vdots \\ \boldsymbol{F}^H_{N_p}\\[0.4em] \boldsymbol{N}^H_1\\ \vdots \\ \boldsymbol{N}^H_{N_p}\\ 
\end{array} \right), \ \ \
\boldsymbol{E}
= \left( \begin{array}{c} 
\boldsymbol{F}^E_1\\ \vdots \\ \boldsymbol{F}^E_{N_p}\\[0.4em] \boldsymbol{N}^E_1\\ \vdots \\ \boldsymbol{N}^E_{N_p}\\ 
\end{array} \right) \nonumber.
\end{eqnarray}
Neglecting direct particle interactions,
the equations of particle motion, i.e., Eqs.~(\ref{e3-2A-1}) and (\ref{e3-2A-2}), are summarized by
\begin{eqnarray}
\boldsymbol{M} \cdot \frac{\mathrm{d}}{\mathrm{d} t} \boldsymbol{U}(t) = \boldsymbol{H}(t) + \boldsymbol{E}(t)
\label{e3-2B-1}
\end{eqnarray}
where the matrix $\boldsymbol{M}$ represents the direct sum of the mass tensor and the moment of inertia tensor.
For spherical particles of equal radii and densities, the matrix $\boldsymbol{M}$ reduces to
\begin{eqnarray}
\boldsymbol{M}
= \left( \begin{array}{cc} 
M\boldsymbol{I}_{3N_p} & \boldsymbol{0}\\ 
\boldsymbol{0} & \frac{2}{5}Ma^2\boldsymbol{I}_{3N_p}\\ 
\end{array} \right)
\label{e3-2B-2}
\end{eqnarray}
where $M$ and $a$ are the mass and the radius, respectively, of each particle.
By performing a Fourier transform on Eq.~(\ref{e3-2B-1}), 
the corresponding equation for the Fourier components with time factor $e^{-i\omega t}$ is obtained as 
\begin{eqnarray}
-i\omega \boldsymbol{M} \cdot \hat{\boldsymbol{U}}(\omega) = \hat{\boldsymbol{H}}(\omega) + \hat{\boldsymbol{E}}(\omega)
\label{e3-2B-3}.
\end{eqnarray}
The following relation holds true for the linearized hydrodynamic equations~\cite{B3-2}:
\begin{eqnarray}
\hat{\boldsymbol{U}}(\omega) = -\hat{\boldsymbol{\mu}}(\omega) \cdot \hat{\boldsymbol{H}}(\omega)
\label{e3-2B-4}
\end{eqnarray}
where $\hat{\boldsymbol{\mu}}$ is the $6N_p \times 6N_p$ mobility matrix.
The mobility matrix depends on the configuration of all of the particles $\{ \boldsymbol{R}_i \}$, so 
we further assume, for simplicity, that the particles do not move significantly over the time scale considered, i.e., the configuration of particles $\{ \boldsymbol{R}_i \}$ is independent of time $t$.
Substituting Eq.~(\ref{e3-2B-4}) into Eq.~(\ref{e3-2B-3}) yields 
an explicit expression for the translational and rotational particle velocities:
\begin{eqnarray}
\hat{\boldsymbol{U}}(\omega) = [\boldsymbol{I} - i\omega \hat{\boldsymbol{\mu}}(\omega) \cdot \boldsymbol{M}]^{-1} 
\cdot \hat{\boldsymbol{\mu}}(\omega) \cdot \hat{\boldsymbol{E}}(\omega)
\label{e3-2B-5}.
\end{eqnarray}
Therefore, for an externally applied force and torque, 
the dynamics of particles in a fluid can be described by the mobility matrix.

The components of the mobility matrix are given by
\begin{eqnarray}
\hat{\boldsymbol{\mu}}(\omega)
= \left( \begin{array}{cc} 
\hat{\boldsymbol{\mu}}^{\rm tt}(\omega) & \hat{\boldsymbol{\mu}}^{\rm tr}(\omega)\\ 
\hat{\boldsymbol{\mu}}^{\rm rt}(\omega) & \hat{\boldsymbol{\mu}}^{\rm rr}(\omega)\\ 
\end{array} \right)
\label{e3-2B-6}
\end{eqnarray}
where $\hat{\boldsymbol{\mu}}^{\alpha \beta}$ is the $3N_p \times 3N_p$ matrix, which is 
composed of the mobility tensors $\hat{\boldsymbol{\mu}}^{\alpha \beta}_{ij}$ for each pairing of particles $i$ and $j$.
The mobility tensors describe the mutual coupling of the translational and rotational motion between particles, 
and the superscript represents the mode of the motion: `t' for translation and `r' for rotation.
The mobility matrix is symmetric as a consequence of the Lorentz reciprocal theorem 
such that the following relations are satisfied~\cite{B3-12}:
\begin{eqnarray}
\hat{\boldsymbol{\mu}}^{\rm tt} = (\hat{\boldsymbol{\mu}}^{\rm tt})^T, \ \ \
\hat{\boldsymbol{\mu}}^{\rm rr} = (\hat{\boldsymbol{\mu}}^{\rm rr})^T, \ \ \
\hat{\boldsymbol{\mu}}^{\rm tr} = (\hat{\boldsymbol{\mu}}^{\rm rt})^T
\label{e3-2B-7}.
\end{eqnarray}

Now, we consider a system of two identical spherical particles in a fluid.
The configuration of the particles is described in Fig.~\ref{f3-1}.
Numbers are assigned to the particles: 1 for the particle on the left and 2 for the particle on the right.
We define a vector $\boldsymbol{R} = \boldsymbol{R}_2 - \boldsymbol{R}_1$
to describe the geometry of this system.
The particle center-to-center distance is denoted by $R = |\boldsymbol{R}|$.
The unit vector along the line of centers is described by $\hat{\boldsymbol{R}} = \boldsymbol{R}/R$.
Due to axisymmetry about the $\hat{\boldsymbol{R}}$ axis, each mobility tensor is described by
at most two scalar functions~\cite{B3-13, B3-14}:
\begin{subequations}
\begin{eqnarray}
\hat{\boldsymbol{\mu}}^{\rm tt}_{ij}(\boldsymbol{R}, \omega) 
= \hat{\mu}^{{\rm tt}\parallel}_{ij}(R, \omega) \boldsymbol{\hat{R}\hat{R}} 
+ \hat{\mu}^{{\rm tt}\perp}_{ij}(R, \omega) (\boldsymbol{I} - \boldsymbol{\hat{R}\hat{R}}) \nonumber, \\
\label{e3-2B-8a}
\end{eqnarray}
\begin{eqnarray}
\hat{\boldsymbol{\mu}}^{\rm rr}_{ij}(\boldsymbol{R}, \omega) 
= \hat{\mu}^{{\rm rr}\parallel}_{ij}(R, \omega) \boldsymbol{\hat{R}\hat{R}} 
+ \hat{\mu}^{{\rm rr}\perp}_{ij}(R, \omega) (\boldsymbol{I} - \boldsymbol{\hat{R}\hat{R}}) \nonumber, \\
\label{e3-2B-8b}
\end{eqnarray}
\begin{eqnarray}
\hat{\boldsymbol{\mu}}^{\rm tr}_{ij}(\boldsymbol{R}, \omega) 
= \hat{\mu}^{{\rm tr}\perp}_{ij}(R, \omega) \boldsymbol{\hat{R}} \times \boldsymbol{I}
\label{e3-2B-8c}.
\end{eqnarray}
\label{e3-2B-8}
\end{subequations}
The superscripts $\parallel$ and $\perp$ denote the directions of motion parallel and perpendicular to the symmetry axis, respectively.
In the direction parallel to the symmetry axis, the translational and rotational motions are decoupled as indicated by Eq.~(\ref{e3-2B-8c}).
Interchanging the particle numbers 1 and 2 corresponds to an inversion of the direction of $\hat{\boldsymbol{R}}$, 
which causes a sign inversion for the translation-rotation cross-mobility tensor $\hat{\boldsymbol{\mu}}^{\rm tr}_{ij}$
without changing the sign of the other cross-mobility tensors, $\hat{\boldsymbol{\mu}}^{\rm tt}_{ij}$ and $\hat{\boldsymbol{\mu}}^{\rm rr}_{ij}$.
The mobility tensors Eqs.~(\ref{e3-2B-8}) within the Oseen approximation are described in the Appendix.

\subsection{Temporal evolution of the flow field}

We first consider a particle dispersion with an incompressible fluid solvent. 
Incompressibility corresponds to an infinite speed of sound in the fluid such that
the solenoidal condition is imposed on the fluid velocity field from Eqs.~(\ref{e3-2A-3}) and (\ref{e3-2A-6}) as
\begin{eqnarray}
\boldsymbol{\nabla} \cdot \boldsymbol{v} = 0
\label{e3-2C-1}.
\end{eqnarray}
The velocity field can be decomposed into two contributions as
\begin{eqnarray}
\boldsymbol{v} = \boldsymbol{w} - \boldsymbol{\nabla}\varphi
\label{e3-2C-2}.
\end{eqnarray}
The vector field $\boldsymbol{w} (\boldsymbol{r}, t)$ is the juxtaposition of the velocity fields in the fluid and particle domains; however, 
the fluid-particle impermeability boundary condition, which demands the continuity of the normal velocity on the boundary, 
is not satisfied.
An irrotational flow field described by the scalar potential $\varphi(\boldsymbol{r}, t)$ is added
so that the velocity field $\boldsymbol{v}$ will satisfy the boundary condition.
Due to the solenoidal condition on the velocity field $\boldsymbol{v}$,
the scalar potential $\varphi$ obeys the Poisson equation:
\begin{eqnarray}
\boldsymbol{\nabla}^2 \varphi = \boldsymbol{\nabla} \cdot \boldsymbol{w}
\label{e3-2C-3}.
\end{eqnarray}
On the right-hand side, $\boldsymbol{\nabla} \cdot \boldsymbol{w}$ is zero except at the fluid-particle boundary at which 
singularities exist.
From Eq.~(\ref{e3-2C-3}), the scalar potential is given by~\cite{B3-15}
\begin{eqnarray}
\varphi(\boldsymbol{r}, t) = -\frac{1}{4\pi} \int \mathrm{d}\boldsymbol{r}' 
\frac{(\boldsymbol{r} - \boldsymbol{r}')}{|\boldsymbol{r} - \boldsymbol{r}'|^3} \cdot \boldsymbol{w}(\boldsymbol{r}', t)
\label{e3-2C-4}.
\end{eqnarray}
If we consider an isolated single particle,
the scalar potential and the corresponding velocity field can be simply described due to axisymmetry:
\begin{eqnarray}
\varphi(\boldsymbol{r}, t) = \frac{\boldsymbol{Q}(t) \cdot \boldsymbol{r}}{r^3}
\label{e3-2C-5},
\end{eqnarray}
\begin{eqnarray}
-\boldsymbol{\nabla}\varphi(\boldsymbol{r}, t) 
= \frac{\boldsymbol{Q}(t) \cdot (3\hat{\boldsymbol{r}}\hat{\boldsymbol{r}} - \boldsymbol{I})}{r^3}
\label{e3-2C-6}
\end{eqnarray}
where $\hat{\boldsymbol{r}} = \boldsymbol{r}/r$ is a unit vector in the radial direction.
The velocity field of Eq.~(\ref{e3-2C-6}) represents a doublet flow,
which corresponds to the electric field generated by an electric dipole.
The vector $\boldsymbol{Q}(t)$ is parallel to the particle velocity 
and depends on time to the same degree as the particle motion relative to the fluid.
The analytical form of $\boldsymbol{Q}(t)$ has been derived previously~\cite{B3-16}, 
and the strength $Q(t) = |\boldsymbol{Q}(t)|$ is time-independent for a neutrally buoyant particle.
The doublet flow is generated instantaneously to satisfy the solenoidal condition;
thus, the doublet flow is interpreted to expand at the infinite speed of sound.

The vector field $\boldsymbol{w}$ represents the shear flow due to viscous diffusion.
When the Reynolds number is sufficiently low, the hydrodynamic equations can be linearized as follows:
\begin{eqnarray}
\rho_0 \frac{\partial \boldsymbol{v}}{\partial t} = -\boldsymbol{\nabla}p + \eta \boldsymbol{\nabla}^2 \boldsymbol{v} + \boldsymbol{f}^R
\label{e3-2C-7}.
\end{eqnarray}
Under the condition of incompressibility, the pressure gradient imposes fluid-particle impermeability on the time-derivative of the velocity field, and
the pressure relates to the scalar potential as
\begin{eqnarray}
p = \rho_0 \frac{\partial \varphi}{\partial t}
\label{e3-2C-8}.
\end{eqnarray}
Therefore, the temporal evolution of the vector field $\boldsymbol{w}$ is given by
\begin{eqnarray}
\rho_0 \frac{\partial \boldsymbol{w}}{\partial t} 
= \eta (\boldsymbol{\nabla}^2 \boldsymbol{w} - \boldsymbol{\nabla}\boldsymbol{\nabla} \cdot \boldsymbol{w}) + \boldsymbol{f}^R
\label{e3-2C-9}.
\end{eqnarray}
Modifying the diffusion term on the right-hand side to be solenoidal 
results in the doublet flow being gradually offset by the diffusion of the vector field $\boldsymbol{w}$.

In short, 
the exertion of an impulsive force on a particle in an incompressible fluid instantaneously generates a doublet flow.
The shear flow then spreads out diffusely, which cancels out the doublet flow.
This picture is validated in the numerical simulation results below.

For a compressible fluid,
a second contribution is added to Eq.~(\ref{e3-2C-2}):
\begin{eqnarray}
\boldsymbol{v} = \boldsymbol{w} - \boldsymbol{\nabla}\psi - \boldsymbol{\nabla}\varphi
\label{e3-2C-10}
\end{eqnarray}
where the scalar potential $\psi(\boldsymbol{r}, t)$ corresponds to the bulk compression flow.
Here, we assume that the dynamics of the fluid can be described by the linearized hydrodynamic equations
\begin{eqnarray}
\frac{\partial p}{\partial t} = - \rho_0 c^2 \boldsymbol{\nabla} \cdot \boldsymbol{v}
\label{e3-2C-11},
\end{eqnarray}
\begin{eqnarray}
\rho_0 \frac{\partial \boldsymbol{v}}{\partial t} = \eta \boldsymbol{\nabla}^2 \boldsymbol{v} 
+ \left(\frac{1}{3}\eta + \eta_v \right) \boldsymbol{\nabla \nabla} \cdot \boldsymbol{v} - \boldsymbol{\nabla}p + \boldsymbol{f}^R \nonumber. \\
\label{e3-2C-12}
\end{eqnarray}
The time evolution equation of the total scalar potential $\varphi' = \varphi + \psi$ is derived from the equations above as
\begin{eqnarray}
\boldsymbol{\nabla}^2 \varphi' - \frac{1}{c^2} \frac{\partial^2 \varphi'}{\partial t^2} 
+ \frac{\eta_l}{\rho_0 c^2}\boldsymbol{\nabla}^2 \frac{\partial \varphi'}{\partial t} \hspace{4.0em} \nonumber\\
= \left( 1 + \frac{\eta_l}{\rho_0 c^2}\boldsymbol{\nabla}^2 \frac{\partial}{\partial t} \right) (\boldsymbol{\nabla} \cdot \boldsymbol{w}) 
\label{e3-2C-13}
\end{eqnarray}
where $\eta_l = (4/3)\eta + \eta_v$ denotes the longitudinal viscosity.
This is a damped wave equation in which the source is given on the right-hand side of the equation.
Therefore, 
a potential flow propagates at a finite speed of sound in a compressible fluid.
In the long-time limit, Eq.~(\ref{e3-2C-13}) reduces to the Poisson equation for an incompressible fluid as given by Eq.~(\ref{e3-2C-3}), 
and the potential flow reduces to an instantaneous doublet flow.

\section{Numerical Results}

Numerical simulations are performed for a three-dimensional box with periodic boundary conditions.
The space is divided into meshes of length $\Delta$, 
which is a unit length.
The units of the other physical quantities are defined by combining $\Delta$ with $\eta = 1$ and $\rho_0 = 1$, 
where $\rho_0$ is the fluid mass density at equilibrium.
The system size is $L_x \times L_y \times L_z = 256 \times 256 \times 256$.
The other parameters are set to $a = 4$, $\xi = 2$, $\rho_p = 1$, $\eta_v = 0$, and $h = 0.05$
where $\rho_p$ is the particle mass density and $h$ is the time increment of a single simulation step.

We consider the system of two identical spherical particles in a fluid 
whose geometry is described in Fig~\ref{f3-1}. 
Both particles have the same density and radius, which also means that they have the same mass and moment of inertia.
We investigate the time-dependence of the velocity for particle 1 following the exertion of an impulsive force at the center of particle 2, 
with changing the center-to-center distance between the particles $R^{\ast} = R/a$.
This cross-relaxation function is a manifestation of the temporal evolution of the hydrodynamic interactions between particles 1 and 2.
The impulsive force is assumed to be sufficiently small such that the Reynolds and Mach numbers of the flow are sufficiently low.  
We set the impulsive force to produce an initial particle Reynolds number of ${\rm Re}_p = 10^{-3}$.
In addition, the displacement of particles is negligible in the present simulations because the particle displacement before stopping, scaled by the particle radius $a$, is comparable to $(\rho_p/\rho_0) {\rm Re}_p$.
Therefore, direct interactions between particles, which include overlap repulsion forces, are not considered because particle collisions do not occur.

The cross-relaxation tensor is introduced as the normalized change in velocity of particle 1:
\begin{eqnarray}
\boldsymbol{V}_1(\boldsymbol{R}_1, t; \boldsymbol{R}_2) = \frac{\boldsymbol{P}_2}{M} \cdot \boldsymbol{\gamma}_{12}(\boldsymbol{R}, t)
\label{e3-3-1},
\end{eqnarray}
\begin{eqnarray}
\boldsymbol{\gamma}_{12}(\boldsymbol{R}, t) 
= \gamma^{\parallel}_{12}(R, t)\hat{\boldsymbol{R}}\hat{\boldsymbol{R}} 
+ \gamma^{\perp}_{12}(R, t)(\boldsymbol{I} - \hat{\boldsymbol{R}}\hat{\boldsymbol{R}})
\label{e3-3-2},
\end{eqnarray}
where $\boldsymbol{P}_2$ is the impulsive force exerted on particle 2 at $t = 0$.
Due to the previously mentioned axisymmetry,
the cross-relaxation tensor is characterized by only two directions of motion, which are
parallel and perpendicular to the center-to-center vector $\boldsymbol{R}$,
and the motions in both directions are decoupled.
From the fluctuation-dissipation theorem, 
the cross-relaxation tensor is equivalent to the velocity cross-correlation function of the fluctuating system:
\begin{eqnarray}
\boldsymbol{\gamma}_{12}(\boldsymbol{R}, t) = 
\frac{M}{k_B T} \langle \boldsymbol{V}_1(\boldsymbol{0}, 0)\boldsymbol{V}_2(\boldsymbol{R}, t) \rangle
\label{e3-3-3},
\end{eqnarray}
where $k_B$ is the Boltzmann constant and $T$ is the thermodynamic temperature.
We examine both the parallel correlation $\gamma^{\parallel}_{12}$ and the perpendicular correlation $\gamma^{\perp}_{12}$ 
by adjusting the direction of the impulsive force $\boldsymbol{P}_2$.
As predicted by the formulation in the preceding section,  
translation-rotation coupling in the particle motion is observed for motion perpendicular to the axis.
However, we focus entirely on translation-translation coupling because the influence of sound propagation on the rotational motion is expected to be small.
We also compare the simulated results with approximate solutions 
wherein the analytical solution for the self-mobilities of isolated single particles and the Oseen approximation for the cross-mobilities are applied (see the Appendix).

\subsection{Incompressible fluid}

\begin{figure}[t]
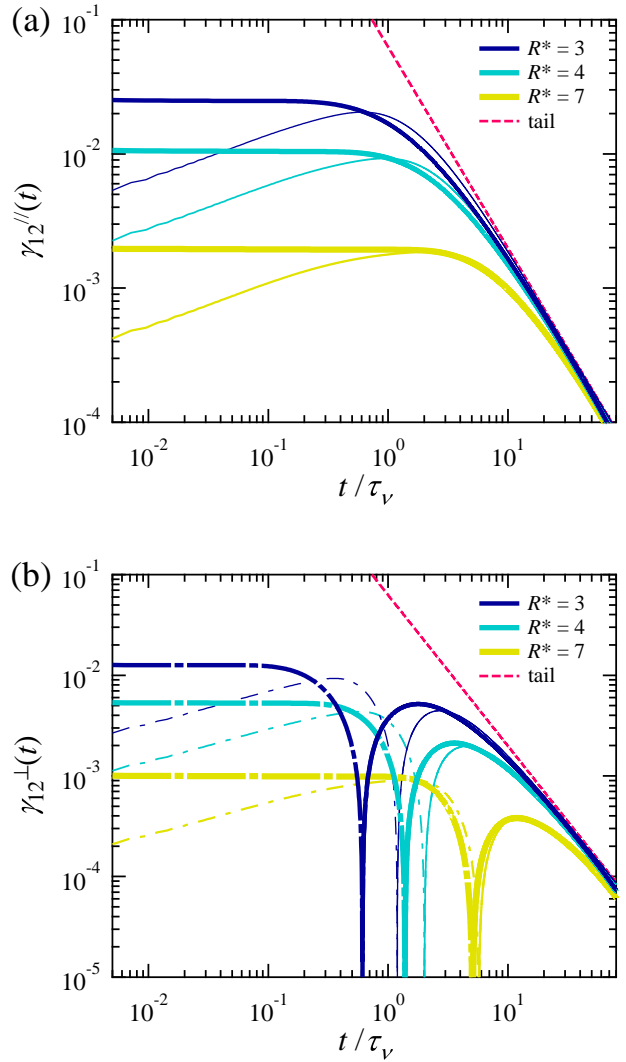

\centering
\includegraphics[width=90mm]{fig3_2a.eps}\hspace{-0.8em}
\includegraphics[width=90mm]{fig3_2b.eps}\\[-1.5em]
\caption{\label{f3-2} Velocity cross-relaxation function in an incompressible fluid in (a) parallel and (b) perpendicular directions relative to the symmetry axis of the particles.
The distances between the two particles are $R^{\ast} =$ 3, 4, and 7.
Simulation results are shown as bold solid lines (positive values) and dashed-dotted lines (negative values).
Solutions within the Oseen approximation are represented by thin solid lines (positive values) and dashed-dotted lines (negative values).
The broken line represents the long-time tail with an algebraic power law decay $B t^{-3/2}$ as given by Eq.~(\ref{e3-3A-1}).
}
\end{figure}

\begin{figure*}[tbp]
\centering
\includegraphics[height=60mm]{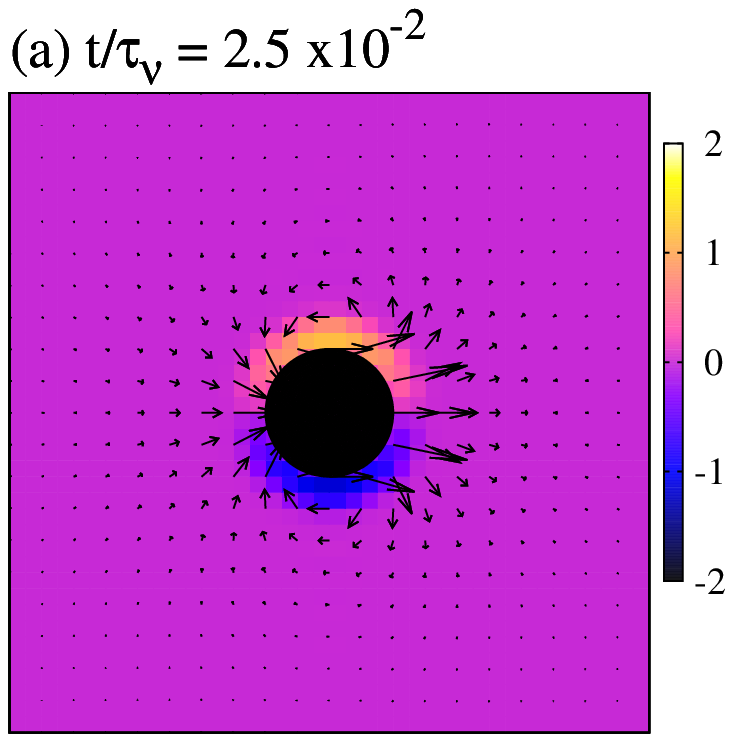}
\includegraphics[height=60mm]{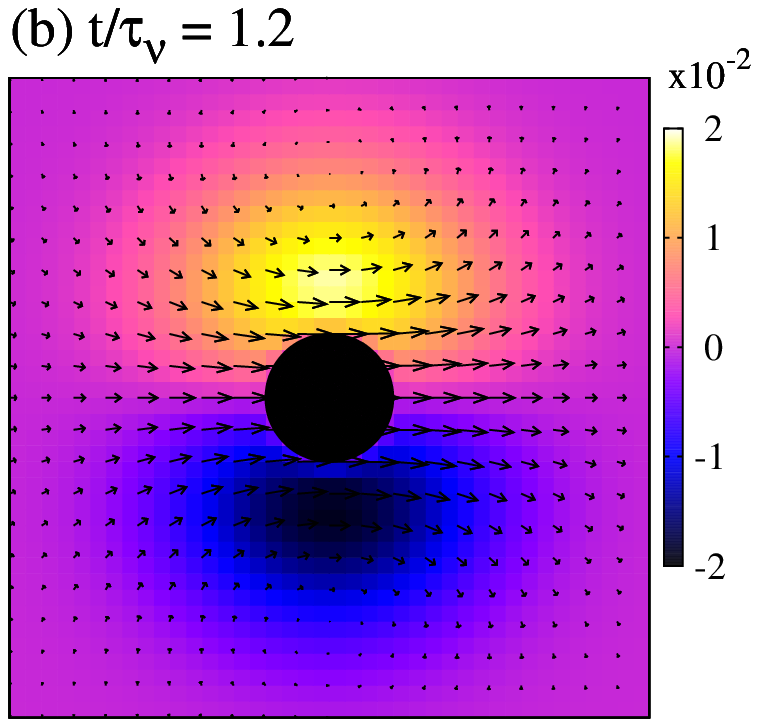}\\[-0.7em]
\caption{\label{f3-3} Temporal evolution of the velocity field around an isolated single particle due to an impulsive force exerted on the particle
in an incompressible fluid.
Cross-sections parallel to the impulsive force direction including the particle center are shown.
The impulsive force is exerted at time $t = 0$, and the simulation results are given at (a) $t/\tau_{\nu} = 2.5\times 10^{-2}$ and (b) $t/\tau_{\nu} = 1.2 $.
The direction of the impulsive force is to the right in the pictures, 
and the particle is represented by a black circle.
The vorticity of the velocity field $\boldsymbol{\nabla} \times \boldsymbol{v}$ is described by a color scale, which goes
from negative (darker) to positive (lighter) vorticity.
The vorticity is normalized by $\tau_{\nu}/{\rm Re}_p$.
}
\end{figure*}

\begin{figure*}[tbp]
\centering
\includegraphics[height=60mm, clip]{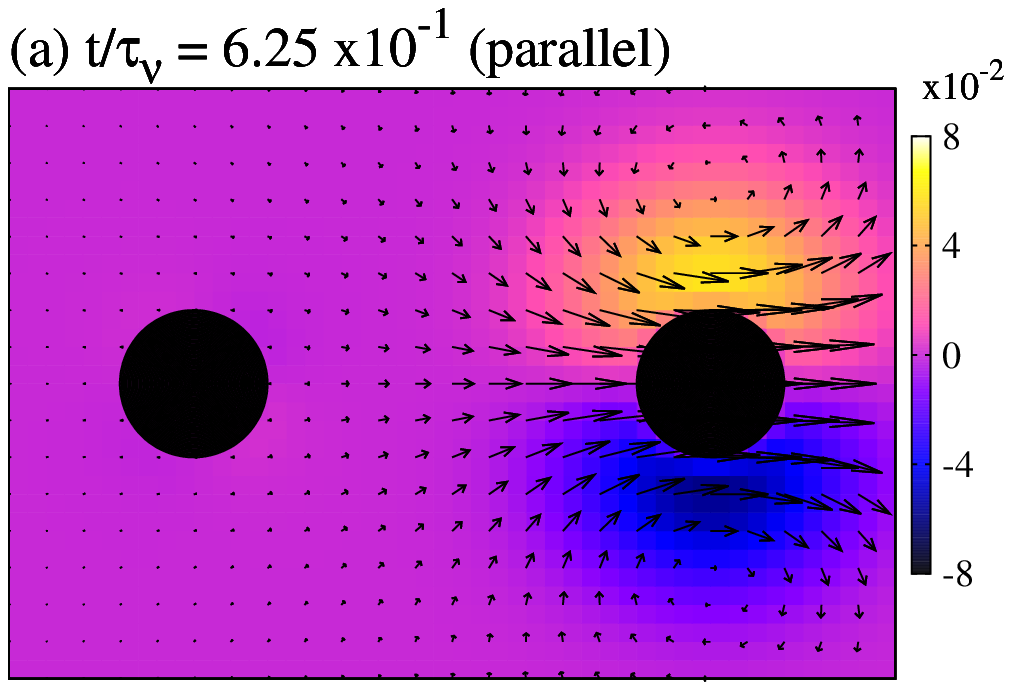}\hspace{-0.4em}
\includegraphics[height=60mm, clip]{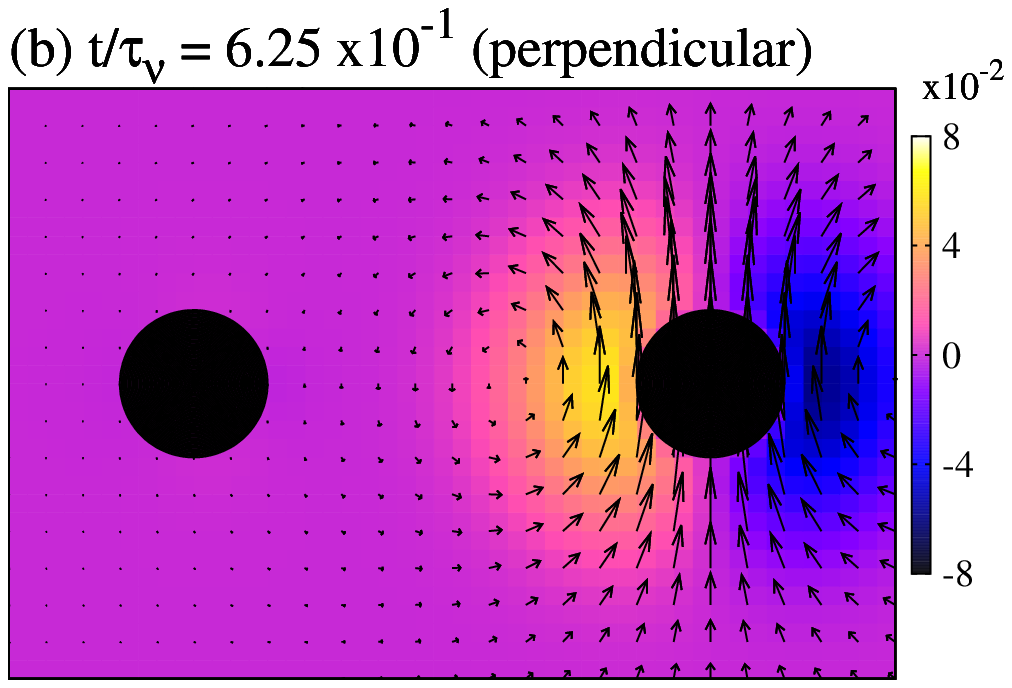}\\[-0.2em]
\includegraphics[height=60mm, clip]{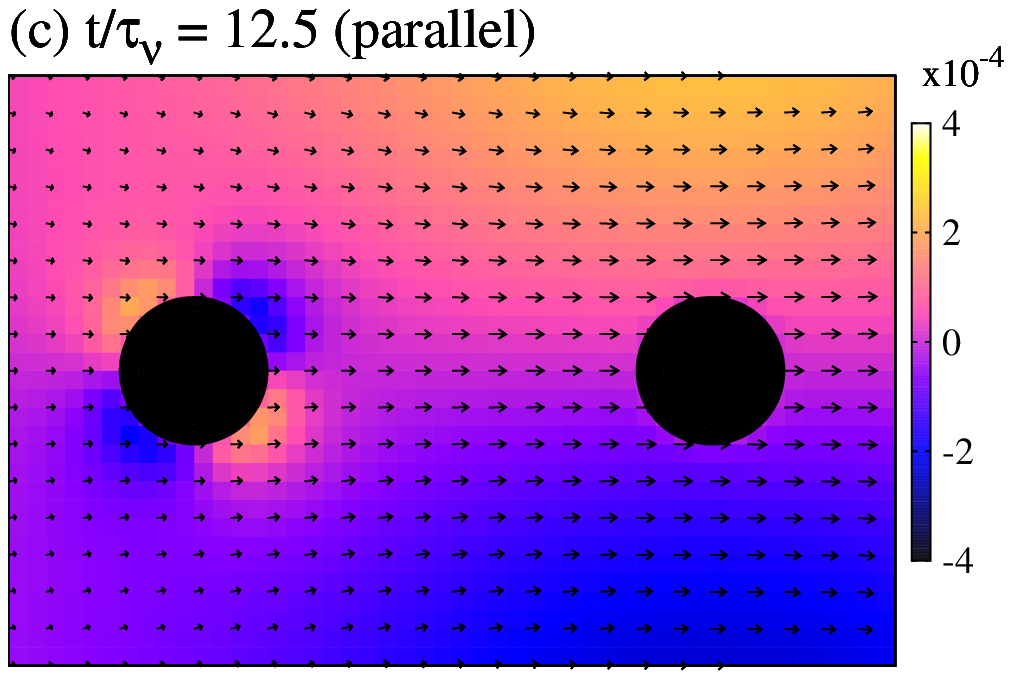}\hspace{-0.4em}
\includegraphics[height=60mm, clip]{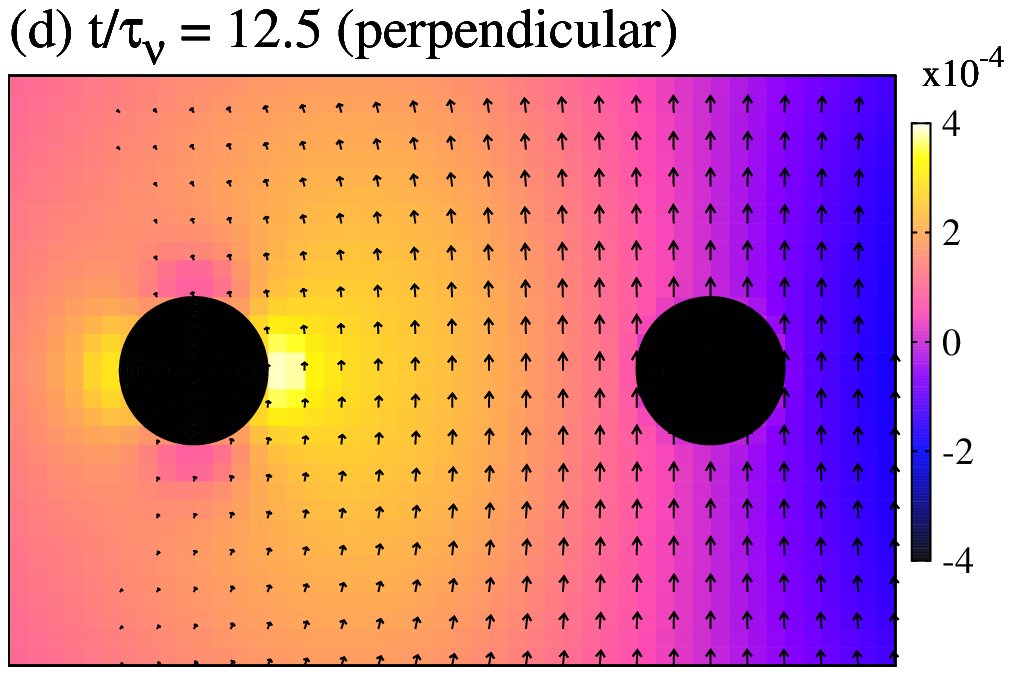}\\[-0.7em]
\caption{\label{f3-4} Temporal evolution of the velocity field around the particles.
Cross-sections including the center-to-center axis between particles are shown.
The center-to-center distance between two particles is $R^{\ast} = 7$.
The impulsive force is exerted at time $t = 0$ on the particle on the right,
and the simulation results are presented at (a,b) $t/\tau_{\nu} = 6.25\times 10^{-1}$ and (c,d) $t/\tau_{\nu} =12.5$.
The direction of the impulsive force is (a,c) to the right or (b,d) upwards. 
The particles are represented by black circles.
The vorticity of the velocity field $\boldsymbol{\nabla} \times \boldsymbol{v}$ is described by a color scale, which ranges
from negative (darker) to positive (lighter) vorticity.
The vorticity is normalized by $\tau_{\nu}/{\rm Re}_p$.
}
\end{figure*}

First, we investigate the temporal evolution of hydrodynamic interactions in an incompressible fluid.
The dynamics of the incompressible fluid is governed by the hydrodynamic equations composed of 
Eqs.~(\ref{e3-2A-4}), (\ref{e3-2A-5}), and (\ref{e3-2C-1}).
The center-to-center distance between particles takes the values of $R^{\ast} =$ 3, 4, and 7.
The simulation results for the velocity cross-relaxation functions are shown in Fig.~\ref{f3-2}.
While the parallel correlation $\gamma^{\parallel}_{12}(t)$ is positive at all times, 
the perpendicular correlation $\gamma^{\perp}_{12}(t)$ is initially negative and subsequently becomes positive.
Comparable results have been reported in experimental studies~\cite{B3-3, B3-4, B3-5}
where the difference between the parallel and perpendicular correlations was attributed to the flow field around the particle~\cite{B3-4, B3-7}.
We discuss the time dependence of the velocity cross-relaxation functions in more detail below.

The temporal evolution of the flow field was described in the previous section.
Here, the velocity fields generated by the impulsive force exerted on a single particle are shown in Fig.~\ref{f3-3}.
These simulation results are obtained for a single particle system with size $L_x \times L_y \times L_z = 128 \times 128 \times 128$.
At early times such as $t/\tau_{\nu} = 2.5 \times 10^{-2}$, the doublet flow is dominant.
The doublet flow is characterized by loop streamlines flared in the direction perpendicular to the particle motion, and 
backflow is observed as described by Eq.~(\ref{e3-2C-5}).
As discussed in the previous section, the doublet flow appears instantaneously and can be interpreted as the propagation of an infinite-speed sound wave.
Conversely, shear flow due to viscous diffusion, 
whose strength is given by the vorticity $\boldsymbol{\nabla} \times \boldsymbol{v}$, 
is only observed in the vicinity of the particle.
At a time $t/\tau_{\nu} = 1.2$, the shear flow has diffused over a large range following Eq.~(\ref{e3-2C-9}); 
therefore, there is a spreading fluid region flowing in the same direction as the particle motion and 
the loop streamlines get away from the particle.

The velocity cross-relaxation function is related to the temporal evolution of the flow field around particle 2.
The doublet flow propagated by sound waves produces an instantaneous velocity correlation between the particles. 
The doublet flow direction shown in Fig.~\ref{f3-3} produces 
a positive parallel correlation and a negative perpendicular correlation.
Almost no change in the cross-relaxation functions occurs in the early stages, which reflects the time-independence of the strength of the doublet flow 
due to the neutrally buoyant particle~\cite{B3-16}.

The reduction of the parallel correlation and the sign inversion of the perpendicular correlation occur at about the same time, 
which depends on the inter-particle distance $R^{\ast}$.
Because shear flow by viscous diffusion can cause such changes in the cross-relaxation functions,
we estimate the time scale on which the shear flow generated by the particle 2 arrives at particle 1 
by the viscous diffusion time scale over the length $L$: $\tau_\nu^{\ast} = L^2/\nu$.
In this case, the characteristic length is the distance between the particle surfaces, $L = R - 2a$; 
thus, for $R^{\ast} =$ 3, 4, and 7, the viscous diffusion time scales are 
$\tau^{\ast}_{\nu}/\tau_{\nu} = (R^{\ast}-2)^2 =$ 1, 4, and 25, respectively.
The time at which the parallel correlation begins to decrease and 
the perpendicular correlation has a sign inversion roughly corresponds to $\tau_\nu^{\ast}$ in each case.

In an incompressible fluid, hydrodynamic interactions are instantly propagated at the infinite speed of sound 
and are subsequently transmitted by viscous diffusion on the time scale $\tau^{\ast}_{\nu}$.
The temporal evolution of the flow fields around two particles separated by $R^{\ast} = 7$ is shown in Fig.~\ref{f3-4}.
At early times in the flow, $t/\tau_{\nu} = 6.25 \times 10^{-1}$, which means that the shear flow has not arrived at particle 1 and that its dynamics are governed by the doublet flow.
Later, at $t/\tau_{\nu} = 12.5$, the shear flow has diffused over time to arrive at particle 1, 
and the motion of particle 1 is governed by the shear flow.
Here, the stresslet flow field is observed in the vicinity of particle 1. 
This flow is generated to maintain particle rigidity against deformation in a shear flow.
In the perpendicular correlation, rotational motion of particle 1 is also observed.

In the final stages of the flow, the momentum of particle 2 has completely diffused away, and  
the particles and the fluid move collectively. 
These dynamics are manifested in a long-time tail with a power law decay $t^{-3/2}$ of the cross-relaxation functions shown in Fig.~\ref{f3-2}, which are given by~\cite{B3-17}
\begin{eqnarray}
\gamma^{\parallel, \perp}_{12}(t) 
= \frac{1}{9 \sqrt{\pi}} \frac{\rho_p}{\rho_0} \left( \frac{\tau_{\nu}}{t} \right)^{3/2} \ \ \ {\rm as} \ \ t \rightarrow \infty
\label{e3-3A-1}.
\end{eqnarray}

Discrepancies between the simulation results and the Oseen approximation are observed especially at $t/\tau_{\nu} \lesssim 3$ as shown in Fig.~\ref{f3-2}.
The approximate cross-relaxation functions clearly change with time in the early stages,
and the time at which the effect of the shear flow becomes apparent is later than in the simulation results.
Particles are assumed to be points in the Oseen approximation;
therefore, it is implied that the particle separation is much larger than the particle radius as $R^{\ast} = R/a \gg 1$,
and the time scale of observation is much longer than that of viscous diffusion over the length of
the particle radius as $t/\tau_{\nu} = t \nu/a^2 \gg 1$.
Neglecting the particle size results in a larger characteristic length $L = R$
and a longer diffusion time scale $\tau^{\ast}_{\nu}/\tau_{\nu} = R^{\ast 2}$.
Consequently, the Oseen approximation can only be accurately applied at long times and large particle separations.
The simulation results shown in Fig.~\ref{f3-2} confirm
that the validity of the Oseen approximation increases with increasing inter-particle distance $R^{\ast}$.
In previous experimental studies, the measurements were conducted over the range for which the Oseen approximation is valid, 
i.e., particle separations $R^{\ast} \gtrsim 5$ and measurement frequencies corresponding to $t/\tau_{\nu} \gtrsim 30$.
Consequently, the experimental results showed good agreement with the Oseen approximation\cite{B3-3, B3-4}.

\subsection{Compressible fluid}

\begin{figure*}[tb]
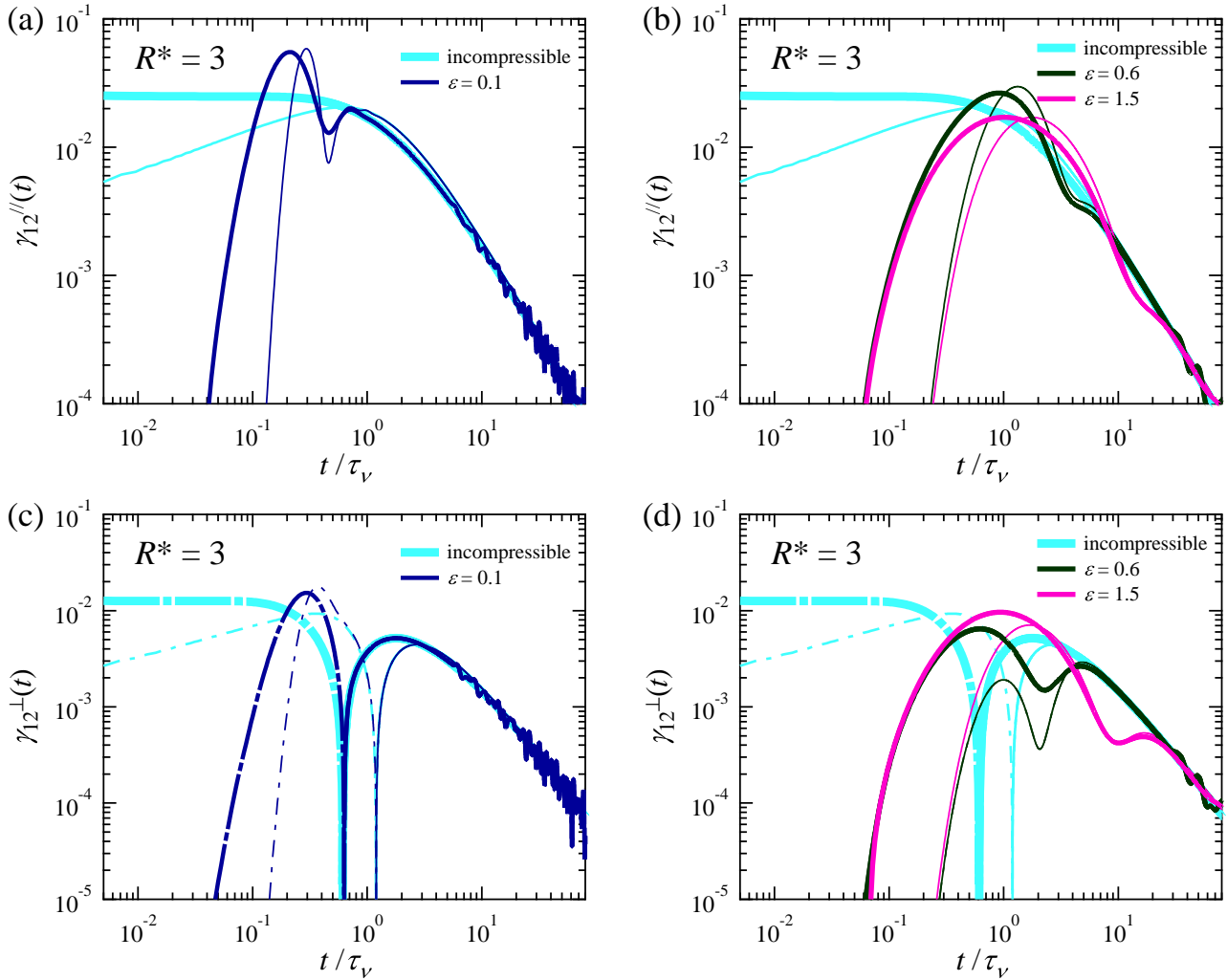

\centering
\includegraphics[width=90mm]{fig3_5a.eps}\hspace{-0.8em}
\includegraphics[width=90mm]{fig3_5b.eps}\\[-1.5em]
\includegraphics[width=90mm]{fig3_5c.eps}\hspace{-0.8em}
\includegraphics[width=90mm]{fig3_5d.eps}\\[-1.5em]
\caption{\label{f3-5} Velocity cross-relaxation function in a compressible fluid in the (a, b) parallel and (c, d) perpendicular directions to the symmetry axis.
The center-to-center distance between two particles is $R^{\ast} = 3$.
The compressibility factor takes the values of $\varepsilon =$ 0.1, 0.6, and 1.5.
Simulation results are shown as bold solid lines (positive values) and dashed-dotted lines (negative values).
Solutions using the Oseen approximation are represented by thin solid lines (positive values) and dashed-dotted lines (negative values).
}
\end{figure*}

\begin{figure*}[t]
\centering
\includegraphics[width=90mm]{fig3_6a.eps}\hspace{-0.8em}
\includegraphics[width=90mm]{fig3_6b.eps}\\[-1.5em]
\includegraphics[width=90mm]{fig3_6c.eps}\hspace{-0.8em}
\includegraphics[width=90mm]{fig3_6d.eps}\\[-1.5em]
\caption{\label{f3-6} Velocity cross-relaxation function in a compressible fluid in the (a, b) parallel and (c, d) perpendicular directions to the symmetry axis.
The center-to-center distance between two particles is $R^{\ast} = 7$.
The compressibility factor takes values of $\varepsilon =$ 0.1, 0.6, and 1.5.
Simulation results are shown as bold solid lines (positive values) and dashed-dotted lines (negative values).
Solutions using the Oseen approximation are represented by thin solid lines (positive values) and dashed-dotted lines (negative values).
}
\end{figure*}

\begin{figure*}[tb]
\centering
\includegraphics[height=60mm, clip]{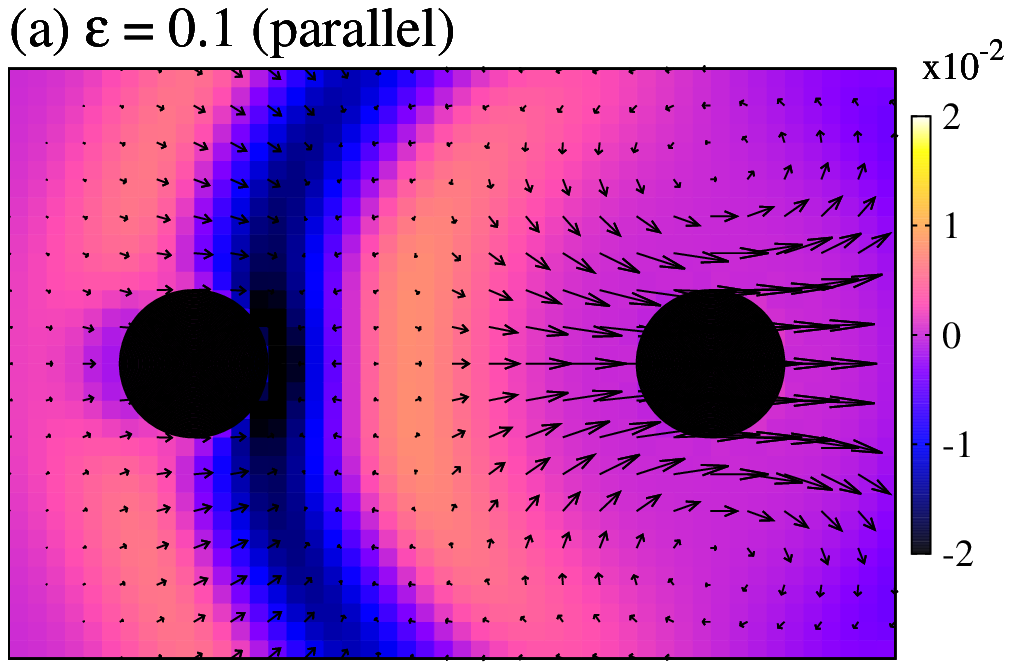}\hspace{-0.4em}
\includegraphics[height=60mm, clip]{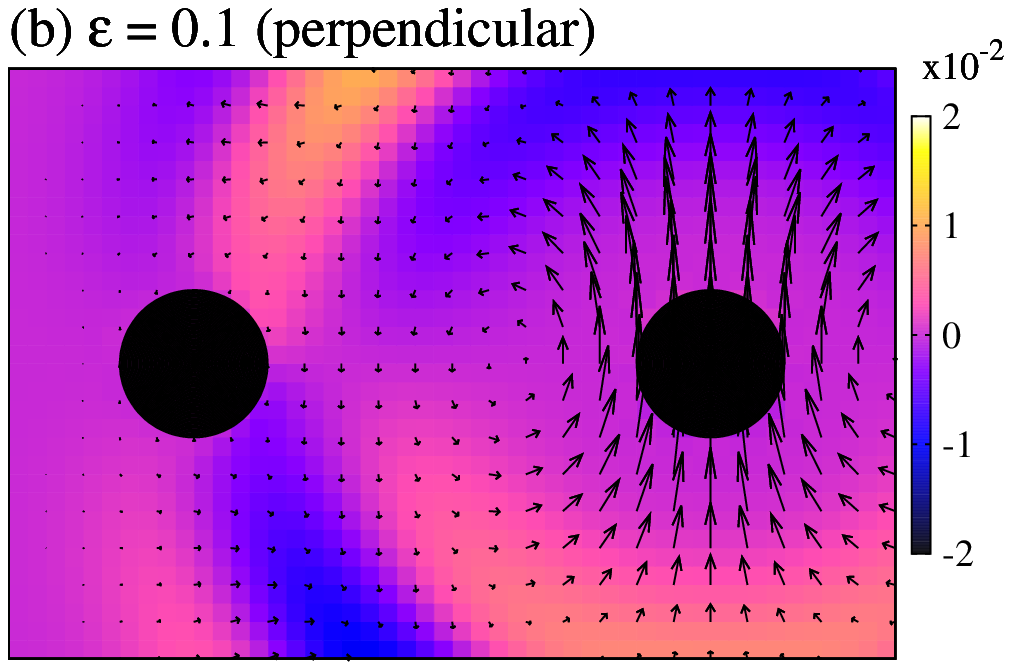}\\[-0.2em]
\includegraphics[height=60mm, clip]{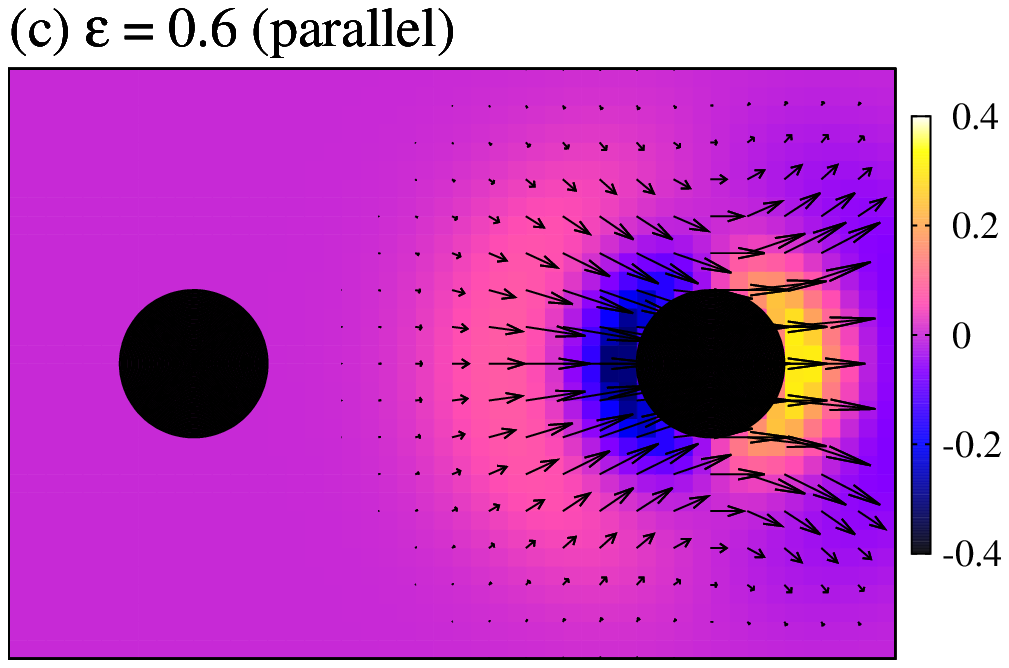}\hspace{-0.4em}
\includegraphics[height=60mm, clip]{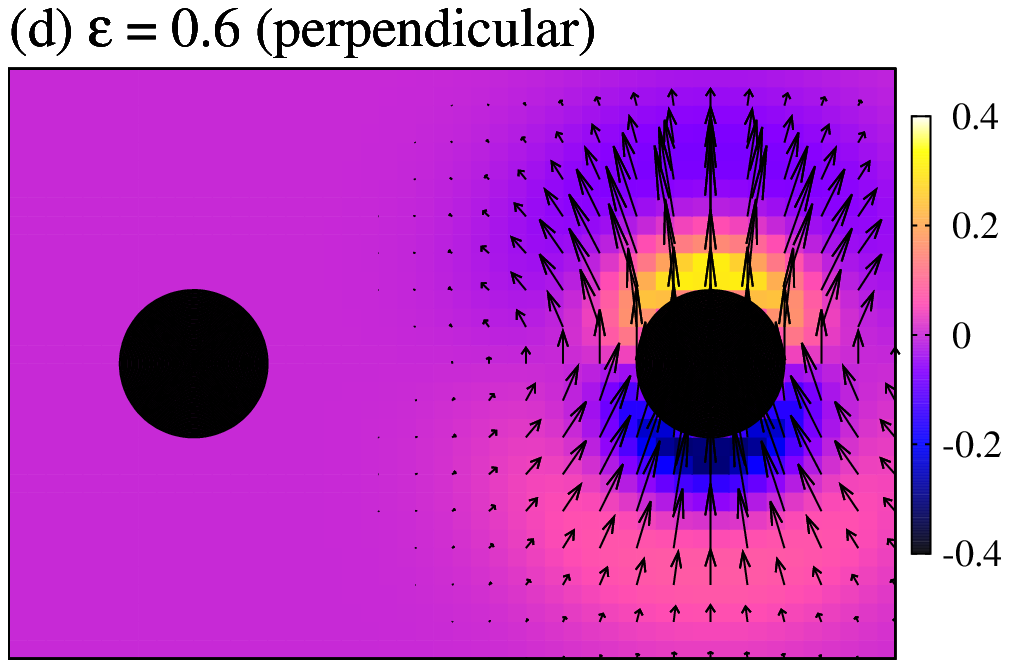}\\[-0.7em]
\caption{\label{f3-7} Velocity field around particles in a compressible fluid.
The simulation results were obtained at $t/\tau_{\nu} = 6.25\times 10^{-1}$ 
for compressibility factors with values (a,b) $\varepsilon = 0.1$ and (c,d) $\varepsilon =0.6$.
The divergence of the velocity field $\boldsymbol{\nabla} \cdot \boldsymbol{v}$ is described by a color scale, which goes
from negative (darker) to positive (lighter) divergence.
The value of divergence is normalized by $\tau_{\nu}/{\rm Re}_p$.
}
\end{figure*}

\begin{figure}[tbp]
\centering
\includegraphics[height=90mm]{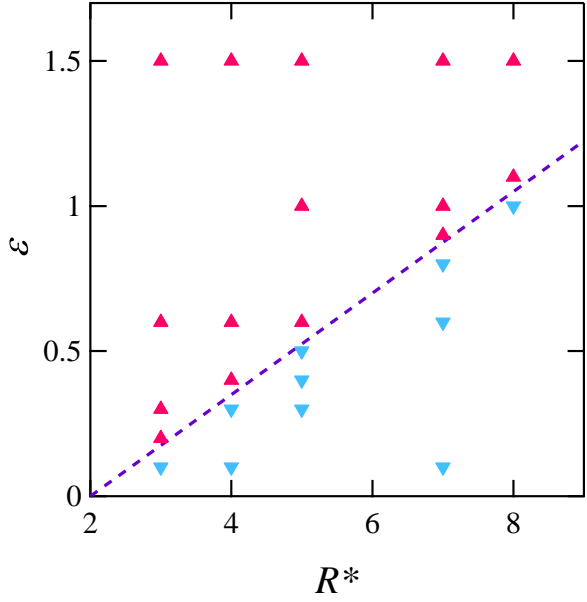}\\[-1.5em]
\caption{\label{f3-8} Classification of calculated cross-relaxation functions for various center-to-center distances of particles and compressibility factors.
The cross-relaxation functions are classified according to the sign of the perpendicular correlation in the first stage: 
negative and positive are represented by downward and upward triangles, respectively.
The broken line represents Eq.~(\ref{e3-3B-1}) with $\varepsilon^{\ast} = 0.175$.
}
\end{figure}

Next, we consider a compressible fluid in which sound propagates at a finite speed.
The velocity cross-relaxation functions for inter-particle distances $R^{\ast} = 3$ and 7 are shown in Figs.~\ref{f3-5} and \ref{f3-6}, respectively.
Due to the periodic boundary conditions imposed on the simulation box, 
a sound pulse generated by the particle motion can continue to affect the simulation results 
after the sound pulse arrives at the edge of the simulation box.
The oscillatory structure of the cross-relaxation functions, especially at $\varepsilon = 0.1$, is one of the striking artifacts of the periodic boundary conditions.

The velocity cross-relaxation functions are zero in the early stages and then suddenly change to non-zero values.
This time lag increases with the inter-particle separation and is likely to be due to 
the sound propagation from particle 2 to particle 1.
As with the diffusion of shear flow, 
we estimate the time scale for sound propagation over a characteristic length $L = R - 2a$ as $\tau^{\ast}_c = L/c$.
For $R^{\ast} =$ 3 and 7, the sound propagation time scales are 
$\tau^{\ast}_c/\tau_{\nu} = (R^{\ast}-2)\varepsilon = \varepsilon$ and $5\varepsilon$, respectively.
The results shown in Figs.~\ref{f3-5} and \ref{f3-6} confirm 
the correspondence between the time scale of sound propagation and the peak of the cross-relaxation functions.
The peak in the cross-relaxation function broadens as sound propagation is dissipated by the longitudinal viscosity according to Eq.~(\ref{e3-2C-13}).
The velocity fields around particles separated by $R^{\ast} = 7$ at $t/\tau_{\nu} = 6.25 \times 10^{-1}$ are shown in Fig.~\ref{f3-7}.
Expanding doublet flows with strengths given by the divergence of the velocity field $\boldsymbol{\nabla} \cdot \boldsymbol{v}$ are observed.
For a small compressibility factor of $\varepsilon = 0.1$, the doublet flow has just arrived at particle 1.
This situation corresponds to a peak in the cross-relaxation functions shown in Fig.~\ref{f3-6}(a,c).
Conversely, for a larger compressibility factor of $\varepsilon = 0.6$, the doublet flow has not yet arrived at particle 1, 
and the cross-relaxation functions are, correspondingly, zero. 

When the fluid compressibility is as small as $\varepsilon = 0.1$,  
the velocity cross-relaxation functions superimpose onto those for the incompressible fluid after a time $\tau^{\ast}_c$.
This behavior only describes hydrodynamic interactions by sound propagation before viscous diffusion effects come into play.
Conversely, for a large compressibility of $\varepsilon = 1.5$,
the behavior of the cross-relaxation functions is considerably different except in the hydrodynamic long-time tail; 
in particular, there is no negative perpendicular correlation.
For a fluid with a medium compressibility such as $\varepsilon = 0.6$,
a negative perpendicular correlation is observed only for the inter-particle separation of $R^{\ast} = 7$.
Because sound propagation and viscous diffusion generate negative and positive perpendicular correlations, respectively, 
a balance of the two time scales $\tau^{\ast}_{\nu}$ and $\tau^{\ast}_c$ is expected to characterize the general behavior of the cross-relaxation function.
Therefore, we define the interactive compressibility factor using a characteristic length $L = R - 2a$ as
\begin{eqnarray}
\varepsilon^{\ast} = \frac{\tau^{\ast}_c}{\tau^{\ast}_{\nu}} = \frac{\varepsilon}{R^{\ast}-2}
\label{e3-3B-1}.
\end{eqnarray}
As the inter-particle distance increases, the two time scales separate to reduce the interactive compressibility factor.
In the present simulations, the interactive compressibility factors for $R^{\ast} =$ 3 and 7 are 
$\varepsilon^{\ast} = \varepsilon$ and $0.2\varepsilon$, respectively.
When the interactive compressibility factor is small, 
hydrodynamic interactions propagating at the speed of sound arrive first at the other particle in about a time $\tau^{\ast}_c$. Following these interactions are 
those propagating by viscous diffusion, which arrive at about a time $\tau^{\ast}_{\nu}$.
The temporal evolution is qualitatively the same as in an incompressible fluid except for the time lag due to sound propagation at a finite speed.
This situation applies to the results where a negative perpendicular correlation is observed in the early stage.
Conversely, the order of arrival of sound propagation and viscous diffusion should be reversed if the interactive compressibility factor is sufficiently large. 
In this situation, the perpendicular correlation is positive from the start due to viscous diffusion, 
and the effect of sound propagation on the hydrodynamic interactions produces a local minimum or even temporary negative values in the cross-correlation function.

In Fig.~\ref{f3-8}, 
the velocity cross-relaxation functions for various center-to-center distances $R^{\ast}$ and compressibility factors $\varepsilon$ are
classified according to the sign of the perpendicular correlation in the first stage.
We can find that the perpendicular correlation is positive from the start when $\varepsilon^{\ast} \geq 0.175$ is satisfied.
In this condition, the hydrodynamic interactions are propagated by viscous diffusion ahead of sound propagation. 

In a compressible fluid, the time condition for the validity of the Oseen approximation is given by $t \gg \max(\tau_{\nu}, \tau_c)$ because there is a finite sound propagation time scale as well as a viscous diffusion time scale.
As with the diffusion of shear flow, 
sound propagation from particle 2 to particle 1 is delayed within the Oseen approximation as shown in Figs.~\ref{f3-5} and \ref{f3-6}.
This delay occurs because the sound propagation time scale is increased within this approximation so that $\tau^{\ast}_c/\tau_{\nu} = R^{\ast}\varepsilon$.
Thus, the discrepancies between the Oseen approximation results with the simulation results can be fairly well-resolved by increasing the particle separation.

\section{Conclusion}

In the present study,
we investigated the temporal evolution of hydrodynamic interactions for a system of two particles 
using SPM to perform direct numerical simulations.
Fluid compressibility was considered in examining temporal evolution by sound propagation.
Hydrodynamic interactions were estimated by the velocity cross-relaxation functions, 
which are equivalent to the velocity cross-correlation functions in a fluctuating system.

In an incompressible fluid, hydrodynamic interactions were observed to propagate instantaneously at the infinite speed of sound followed by
subsequent temporal evolution by viscous diffusion. 
Theoretical analysis showed that  
a doublet flow and a shear flow are generated by sound propagation and viscous diffusion, respectively.
Because the cross-relaxation function reflects the characteristics of each flow field, 
sound propagation and viscous diffusion are associated with negative and positive perpendicular correlations between the particle velocities, respectively.
Therefore, the time at which the behavior of the cross-relaxation function changes is related to 
the time scale of viscous diffusion over the particle separation $\tau^{\ast}_{\nu}$, 
which is the only time scale characterizing the temporal evolution of hydrodynamic interactions in an incompressible fluid. 

In a compressible fluid, sound propagates between particles in a finite time that scales as $\tau^{\ast}_c$.
The effect of the order-of-magnitude relationship between the two time scales $\tau^{\ast}_{\nu}$ and $\tau^{\ast}_c$ on the cross-relaxation function was observed.
An interactive compressibility factor $\varepsilon^{\ast}$ was defined as the ratio of the two time scales given by Eq.~(\ref{e3-3B-1}).
In our simulation results, 
a reversal of the order of arrival of sound propagation and viscous diffusion at the other particle was observed at $\varepsilon^{\ast} \geq 0.175$; in this case,  
the perpendicular correlation is positive from the start and 
hydrodynamic interactions were largely governed by viscous diffusion.

The temporal evolution of hydrodynamic interactions does not qualitatively change from that for an incompressible fluid 
as long as the interactive compressibility factor is small.
Only when the interactive compressibility factor is sufficiently large can 
differences in the temporal evolution of hydrodynamic interactions be expected.
Such a situation could be realized in a highly viscous fluid, and
experimental validation of the results presented here is desirable.

\section*{Acknowledgements}
This work was supported by KAKENHI 23244087 and 
the JSPS Core-to-Core Program ``International research network for non-equilibrium dynamics of soft matter.''

\section*{Appendix: Mobility Matrix in the Oseen Approximation}
\renewcommand{\theequation}{A\arabic{equation}}
\setcounter{equation}{0}
Hydrodynamic interactions among particles in low Reynolds number flows are described by the mobility matrix discussed in $\S$2.2.
However, there is no known closed-form solution even for a two-particle system.
In this section, we introduce an approximate analytical form for the mobility tensors.

For low Reynolds number flows, the hydrodynamic equations can be linearized as given by Eqs.~(\ref{e3-2C-11}) and (\ref{e3-2C-12}), 
and the corresponding equations for the Fourier components with the time factor $e^{-i\omega t}$ are obtained as
\begin{eqnarray}
-i\omega \hat{p}_{\omega} + \rho_0 c^2 \boldsymbol{\nabla} \cdot \hat{\boldsymbol{v}}_{\omega} = 0
\label{e3-A-1},
\end{eqnarray}
\begin{eqnarray}
-i\omega \hat{\boldsymbol{v}}_{\omega} = \eta \boldsymbol{\nabla}^2 \hat{\boldsymbol{v}}_{\omega} 
+ \left(\frac{1}{3}\eta + \eta_v \right) \boldsymbol{\nabla \nabla} \cdot \hat{\boldsymbol{v}}_{\omega} 
- \boldsymbol{\nabla}\hat{p}_{\omega} + \hat{\boldsymbol{f}}^R_{\omega}. \nonumber \\
\label{e3-A-2}
\end{eqnarray}
The fluid velocity field generated by the body force $\boldsymbol{f}^R$ is expressed as
\begin{eqnarray}
\hat{\boldsymbol{v}}_{\omega}(\boldsymbol{r}) = 
\int \mathrm{d}\boldsymbol{r}' \boldsymbol{\hat{G}}(\boldsymbol{r}-\boldsymbol{r}', \omega) \cdot \hat{\boldsymbol{f}}^R_{\omega}(\boldsymbol{r}')
\label{e3-A-3},
\end{eqnarray}
where the Green's function is given by~\cite{B3-18,B3-19}
\begin{eqnarray}
\boldsymbol{\hat{G}}(\boldsymbol{r}, \omega) = \frac{1}{4 \pi \eta} 
\left( \frac{e^{-\alpha r}}{r} \boldsymbol{I} + \alpha^{-2} \boldsymbol{\nabla \nabla} \frac{e^{i\mu r} - e^{-\alpha r}}{r} \right), \nonumber \\
\label{e3-A-4}
\end{eqnarray}
with 
\begin{eqnarray}
\alpha = (-i\omega \rho_0 / \eta)^{1/2},\ \ \ \ \mu = \omega / \tilde{c}
\label{e3-A-5},
\end{eqnarray}
and
\begin{eqnarray}
\tilde{c} = c \left[ 1 - \frac{i \omega}{\rho_0 c^2} \left( \frac{4}{3} \eta + \eta_v \right) \right]^{1/2}
\label{e3-A-6}.
\end{eqnarray}
For an isolated single spherical particle in a fluid
with a body force constraint $\boldsymbol{f}^R$ to satisfy the condition of particle rigidity,
the self-mobilities can be analytically derived~\cite{B3-18,B3-20,B3-21} as follows:
\begin{eqnarray}
\hat{\mu}^{\rm{tt}}_{11}(\omega) = \mu^{\rm{t}}_0 \frac{9}{2x^2} \frac{2x^2 (1-iy) - (1+x)y^2 - x^2 y^2}{(1+x)(9-9iy-2y^2) + x^2 (1-iy)}, \nonumber \\
\label{e3-A-7}
\end{eqnarray}
\begin{eqnarray}
\hat{\mu}^{\rm{rr}}_{11}(\omega) = \mu^{\rm{r}}_0 \frac{3(1 + x)}{3 + 3x + x^2}
\label{e3-A-8},
\end{eqnarray}
where $x = \alpha a$, $y = \mu a$, $\mu^{\rm{t}}_0 = (6\pi \eta a)^{-1}$, and $\mu^{\rm{r}}_0 = (8\pi \eta a^3)^{-1}$.
When the speed of sound is assumed to be infinite, i.e., $\mu = 0$, the solution for an incompressible fluid is obtained.
The translation-rotation coupling does not appear in the self-mobility, i.e., $\hat{\mu}^{\rm{rt}}_{11} = 0$.

To calculate the cross-mobilities, we introduce the Oseen approximation in which the particles are regarded as points.
In the Oseen approximation, a particle exerts a Stokeslet (point force);
therefore, according to Eq.~(\ref{e3-A-3}), 
the translation-translation cross-mobility is the Green's function itself:
\begin{eqnarray}
\hat{\boldsymbol{\mu}}^{\rm{tt}}_{12}(\boldsymbol{R}, \omega) = \boldsymbol{\hat{G}}(\boldsymbol{R}, \omega)
\label{e3-A-9}.
\end{eqnarray}
The components of the parallel and perpendicular directions are explicitly given by
\begin{subequations}
\begin{eqnarray}
\hat{\mu}^{\rm{tt}\parallel}_{12}(R, \omega) = \mu^{\rm{t}}_0 \frac{3a}{2\alpha^2 R^3} [(2 - 2i\mu R - \mu^2 R^2)e^{i\mu R} \nonumber \\
 - (2 + 2\alpha R)e^{-\alpha R}], \hspace{1.0em}
\label{e3-A-10a}
\end{eqnarray}
\begin{eqnarray}
\hat{\mu}^{\rm{tt}\perp}_{12}(R, \omega) = \mu^{\rm{t}}_0 \frac{3a}{2\alpha^2 R^3} [(-1 + i\mu R)e^{i\mu R} \nonumber \\
 + (1 + \alpha R + \alpha^2 R^2)e^{-\alpha R}]
\label{e3-A-10b}.
\end{eqnarray}
\label{e3-A-10}
\end{subequations}
The Stokeslet also generates a rotational motion with an angular velocity of $(1/2)\boldsymbol{\nabla} \times \hat{\boldsymbol{v}}_{\omega}$;
therefore, the translation-rotation cross-mobility is given by
\begin{eqnarray}
\hat{\boldsymbol{\mu}}^{\rm{rt}}_{12}(\boldsymbol{R}, \omega) 
= -\frac{1}{2} \boldsymbol{\nabla} \times \boldsymbol{\hat{G}}(\boldsymbol{R}, \omega)
\label{e3-A-11},
\end{eqnarray}
which has a perpendicular-only component 
\begin{eqnarray}
\hat{\mu}^{\rm{rt}\perp}_{12}(R, \omega) = \mu^{\rm{r}}_0 \frac{a^3}{R^2} (1 + \alpha R)e^{-\alpha R}
\label{e3-A-12}.
\end{eqnarray}
There are no effects from the fluid compressibility.
This result can also be derived by considering the velocity field generated by a rotlet (point torque)~\cite{B3-22}.
The rotation-rotation cross-mobility is obtained from the angular velocity generated by a rotlet:
\begin{eqnarray}
\hat{\boldsymbol{\mu}}^{\rm{rr}}_{12}(\boldsymbol{R}, \omega) 
= \frac{1}{4} \boldsymbol{\nabla} \times \boldsymbol{\nabla} \times \boldsymbol{\hat{G}}(\boldsymbol{R}, \omega)
\label{e3-A-13}.
\end{eqnarray}
Then, each component is given by
\begin{subequations}
\begin{eqnarray}
\hat{\mu}^{\rm{rr}\parallel}_{12}(R, \omega) = \mu^{\rm{r}}_0 \frac{a^3}{R^3} (1 + \alpha R)e^{-\alpha R}
\label{e3-A-14a},
\end{eqnarray}
\begin{eqnarray}
\hat{\mu}^{\rm{rr}\perp}_{12}(R, \omega) = -\mu^{\rm{r}}_0 \frac{a^3}{2R^3} (1 + \alpha R + \alpha^2 R^2)e^{-\alpha R}. \nonumber \\
\label{e3-A-14b}
\end{eqnarray}
\label{e3-A-14}
\end{subequations}

Using the analytical solutions for the self-mobilities of isolated single particles and the Oseen approximation for the cross-mobilities,
the velocity cross-relaxation functions can be approximated by Eq.~(\ref{e3-2B-5}).

\nocite{*}

\bibliography{aipsamp}

\end{document}